\documentclass[aps,twocolumn,superscriptaddress]{revtex4}
\usepackage{amsfonts}
\usepackage{amsmath}
\usepackage{amssymb}
\usepackage{graphicx}
\usepackage{float}
\usepackage{color}
\begin{document}
\title{Activation of a supercooled liquid confined in a nanopore. }

\author{Felix Mercier}
\affiliation{Laboratoire de Photonique d'Angers EA 4464, Universit\' e d'Angers, Physics Department,  2 Bd Lavoisier, 49045 Angers, France}

\author{Gaetan Delhaye}
\affiliation{Laboratoire de Photonique d'Angers EA 4464, Universit\' e d'Angers, Physics Department,  2 Bd Lavoisier, 49045 Angers, France}

\author{Victor Teboul}
\email{victor.teboul@univ-angers.fr}
\affiliation{Laboratoire de Photonique d'Angers EA 4464, Universit\' e d'Angers, Physics Department,  2 Bd Lavoisier, 49045 Angers, France}


\begin{abstract}
It is well stablished that confinement of supercooled liquids in nano-pores induces various effects as a strong modification of the dynamics and a layering of the local structure.
 In this work we raise the issue as how  these confinement effects are modified when the liquid is out of equilibrium. 
To answer that question, we use molecular dynamics simulations to investigate the effect of confinement on a supercooled liquid activated by the periodic folding of a molecular motor.
We find that the motor's opening angle controls the activation of the medium and use that result to study the effect of different activations on the confined supercooled liquid.
We observe an increase of the activation effect on the dynamics when the medium is confined.
We find that the confinement slowing down dependence with the pore radius depends on the activation of the liquid. 
We argue that these findings result from a modification of dynamic correlation lengths with activation. 
In this picture the activation permits to control the liquid correlation length and dynamics inside the pore.
Studying the local structure we observe a modification of the layering organization induced by the activation.
We also find that the mobility inside the pore depends on the layers, being larger where the local density is small.

\end{abstract}

\maketitle
\section{ Introduction}

Liquids and solids constituted or comprising molecules like molecular motors that are able to move by themselves are called active\cite{active1,active5,active-confi}.
An example is cells of living biological organisms that are constituted of active soft materials.
Active matter opens a window to the statistical physics of non equilibrium systems and to biological physics making it a  fascinating new domain of research that develops rapidly\cite{active1,active2,active3,active4,active5,active6,active7,active8,active9,active10,active11,active12,active13,active14,active15,active16,active17,active18,active-confi}.
The different statistical physics associated with non equilibrium systems, induces a number of new phenomena, as for example non-equilibrium phase transitions\cite{active4}.

\begin{figure}
\centering
\includegraphics[height=11.5 cm]{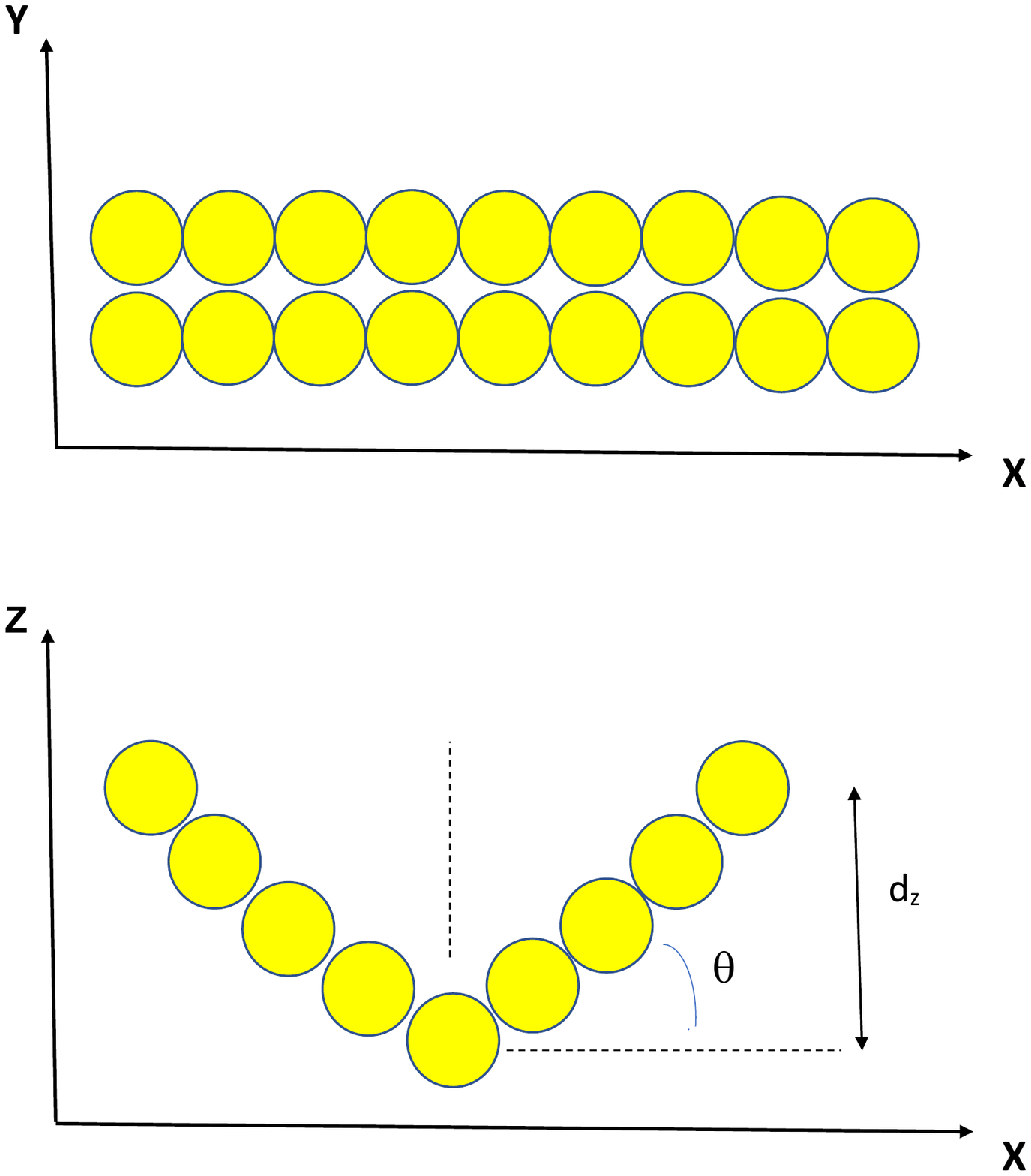}
\caption{(color online) Picture of the folded and unfolded motor molecule. Through this work we use the parameter $d_{z}$  to tune and quantify the activation of the medium by the motor's folding. 
We validate that choice at posteriori in the excitation density section.}
\label{f0}
\end{figure}

\begin{figure}
\centering
\includegraphics[height=8. cm]{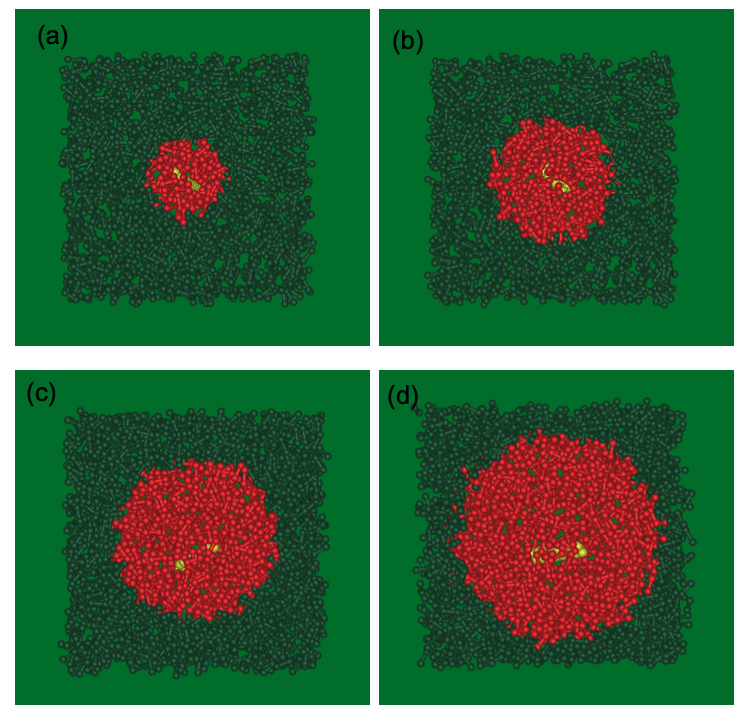}

\caption{(color online) Snapshots of simulations for pores of radii (a) $R=5$\AA, (b) $R=8$\AA, (c) $R=10$\AA, and (d) $R=12.5$\AA. The colors are arbitrary.  $T=500K$. 
The simulations use $8$ pores of radii varying from $5$ to $12.5$\AA\ separated by steps of $1$ \AA\ ($1.5$ \AA\ for the last step).
}
\label{f01}
\end{figure}

\begin{figure}
\centering
\includegraphics[height=6. cm]{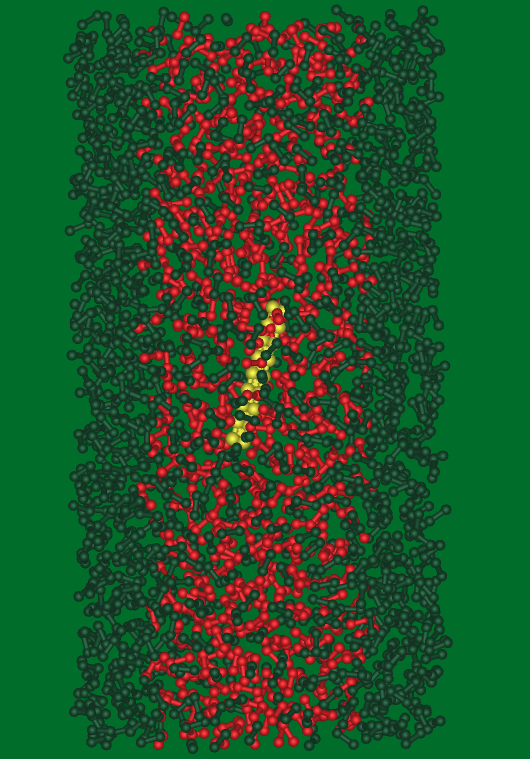}

\caption{(color online) Snapshot of a simulation observed from the side to show the motor. The pore radius is $R=10$\AA. $T=500 K$.}
\label{f02}
\end{figure}

When liquids are cooled fast enough to remain liquid below their melting temperature $T_{m}$, they are subject to a large increase of their viscosity. The viscosity below $T_{m}$ then increases at least exponentially when the temperature drops, leading eventually to a glass at the glass-transition temperature $T_{g}$. That dramatic increase of the viscosity and the associated decrease of most dynamical properties like the diffusion coefficient, strangely appears without significant structural modifications of the liquid.
The nature of that transition from the liquid to the glass, known as the long standing glass-transition problem, is still puzzling scientists\cite{anderson,gt0,gt1,gt2,dh0,c3,c4,md16}.
As phase transitions are usually associated with the divergence of  correlation lengths and the apparition of cooperative mechanisms, increasing correlation lengths  have been searched extensively in supercooled liquids. Confinement of the liquid inside nano-pores has been seen as a way to cutoff the correlation lengths and obtain indirect informations on them by studying the associated dynamical modifications in the liquid. For that fondamental reason as well as multiple practical applications, confinement has been the subject of a number of studies and is still a subject of active research\cite{conff1,conff2,conff3,conff4,conff5,conff6,conf-1,conf0,conf1,conf2,conf3,conf4,conf5,conf6,confine,conf7,active-confi}. 
Indeed, in supercooled liquids, confinement induces various effects.
As suggested by the theoretical perspective described above, confinement in most cases induces a slowing down of the dynamics reminiscent of the slowing down induced by a temperature decrease.
It also induces a structural organization called layering due to the fixation of the local structure in front of the immobile walls of the pore.
In this work we raise the question to how  the confinement effects are modified for active matter.
In order to answer that question, we use a simple butterfly-like flat molecular motor\cite{motoro1,motoro2,motoro3,motoro4,motoro5,motoro6,motoro7,motoro8,motoro9,motoro10,motoro11} (Figure \ref{f0}), to activate with its periodic folding the supercooled liquid surrounding it, confined in nano-pores of various radii (Figures \ref{f01} and \ref{f02}). Experimentally our motor can be seen as a simplification of the azobenzene molecule or derivatives, which undergoes a photo-isomerization process\cite{az1,az2,az3,az4,az5,az6,az7,az8,az9,cage,md16,az10} when subject to a light stimulus.
 We study the effect of different activations induced by different opening angles of our motor (Figure \ref{f0}).
We observe an increase of the activation effect on the dynamics when the medium is confined.
Studying the local structure we observe a modification of the layering organization induced by the activation.
 We also find that the motility is dependent on the layers, being larger where the local density is small.

\section{Calculation}

In addition to theoretical and experimental methods, simulations in particular  molecular dynamics and Monte Carlo simulations\cite{md1,md2,md2b,md4} together with model systems\cite{ms1,ms2,ms3,ms4,ms5}  are now widely used to unravel unsolved problems in condensed matter and complex systems physics\cite{keys,md3,md4b,md6,md7,md8,md9,ee2,md11,md12,md13,md14,md15,ee1,finite1,u1}. 
 The reader will find details on our simulation procedure in  previous papers\cite{pccp,prefold,ariane}, however for convenience we will resume it here.
Our simulations use  one motor molecule  (see Figure \ref{f0} for its description) imbedded inside a medium constituted of $1760$ linear molecules, in a parallelepiped box  $31.1$ \AA\ wide and $62.2$ \AA\ high. 
After aging the simulation box during $10 ns$ at the temperature of studies, we create the pore by suddenly freezing molecules located at a distance larger than $R$ from the center axis of our simulation box.
This procedure is intended to minimize the structure modification of the liquid by the confinement.
After their sudden freezing, the glassy molecules of the pores are no more allowed to move, leading to pure elastic interactions with the confined liquid.
Consequently the liquid temperature is not modified by the interactions with the pore walls.
After that procedure, as the dynamics changes with the confinement, the liquid is again aged during $10 ns$ before the beginning of any study.

We integrate the equations of motion using the Gear algorithm with a quaternion decomposition\cite{md1}  and a time step $\Delta t=10^{-15} s$. 
When the motor is active our simulations are out of equilibrium, because the motor's folding periodically release energy into the system.
However our system is in a steady state and does not age as explained below.
For that purpose we use the Berendsen thermostat\cite{berendsen} to evacuate from the system the energy created by the motor's folding. 
Notice that the thermostat applies to the medium's molecules but not to the motor. Thus the motor's cooling results only from the molecular interactions with the medium's molecules.
We use the NVT canonic thermodynamic ensemble as approximated by Berendsen  thermostat   (see ref.\cite{finite2} for an evaluation of the effect of the thermostat on our calculations), and periodic boundary conditions. 
 The molecules of the medium (host)\cite{ariane} are constituted of two rigidly bonded atoms ($i=1, 2$) at the fixed interatomic distance  {\color{black} $l_{h}$}$=1.73 $\AA$ $. These atoms interact with atoms of other molecules with the following Lennard-Jones potentials:
\begin{equation}
V_{ij}=4\epsilon_{ij}((\sigma_{ij}/r)^{12} -(\sigma_{ij}/r)^{6})   \label{e1}
\end{equation}
with the parameters\cite{ariane}: $\epsilon_{11}= \epsilon_{12}=0.5 KJ/mol$, $\epsilon_{22}= 0.4 KJ/mol$,  $\sigma_{11}= \sigma_{12}=3.45$\AA, $\sigma_{22}=3.28$\AA.
The mass of the motor is $M=540 g/mole$ (constituted of $18$ atoms, each one of mass $30g/mole$) and the mass of the host molecule is $m=80 g/mole$ ($2$ atoms with a mass of $40g/mole$ each).
We model the motor with $18$ atoms in a rectangular shape constituted of two rows of  $9$ rigidly bonded atoms. 
The width of the swimmer is $L_{s}=4.4$\AA\ and its length $l_{s}=15.4$\AA. The length of the host molecule is $l_{h}=5.09$\AA\ and its width $L_{h}=3.37$\AA.
The motor's atoms interact with the medium's atoms using mixing rules and a Lennard-Jones interatomic potential on each atom of the motor, defined by the parameters: 
$\epsilon_{33}= 1.88 KJ/mol$,  $\sigma_{33}=3.405$\AA. 
We use the following mixing rules \cite{mix1,mix2}: 
\begin{equation}
\epsilon_{ij}=(\epsilon_{ii} . \epsilon_{jj})^{0.5}  ; 									
\sigma_{ij}=(\sigma_{ii} . \sigma_{jj})^{0.5}   \label{e2}
\end{equation}
  for the interactions between the motor and the host atoms. 
Our medium is a fragile liquid\cite{fragile1,fragile2}  that falls out of equilibrium in bulk simulations below $T=340 K$, i.e. $T=340$ K is the smallest temperature for which we can equilibrate the bulk liquid when the motor is not active. 
Consequently  the bulk medium above that temperature behaves as a viscous supercooled liquid in our simulations while below it behaves as a solid (as $t_{simulation}<\tau_{\alpha}$). 
Notice however that when active, the motor's motions induce a fluidization of the medium around it\cite{md16,flu1,flu2,flu3,flu4,carry,flu5,rate}, an effect that has been experimentally demonstrated\cite{flu1,flu2,flu3,flu4,flu5} with  azobenzene  photo-isomerizing molecules embedded in soft matter. 
We evaluate the glass transition temperature $T_{g}$ in the bulk to be slightly smaller  $T_{g} \approx 250 K$.
Notice that as they are modeled with Lennard-Jones atoms, the host and motor potentials are quite versatile.
Due to that property, a shift in the parameters $\epsilon$ will shift all the temperatures by the same amount, including the glass-transition temperature and the melting temperature of the material.
Each folding is modeled as continuous, using a constant quaternion variation, with a folding time $\tau_{f}=0.3 ps$.
The total cycle period is constant and fixed at $\tau_{p}=400 ps$.
The folded and unfolded part of the cycle have the same duration equal to $\tau_{p}/2=200 ps$ including the $0.3 ps$ folding or unfolding.
Our system, while out of equilibrium, is in a steady state and is not aging. That behavior is obtained because the energy released by the motor into the medium is small enough and the time lapse between two stimuli large enough for the system to relax before a new stimuli appears.
In other words we are in the linear response regime\cite{pccp}.
Through this work we use the mean square displacement to  obtain informations on the molecules mobility and diffusion inside the pore. 
The mean square displacement is defined as\cite{md1}:\\

\begin{equation}
\displaystyle{<r^{2}(t)>= {1\over N.N_{t_{0}}} \sum_{i,t_{0}} \mid{{\bf r}_{i}(t+t_{0})-{\bf r}_{i}(t_{0})} }\mid^{2}   \label{e130}   
\end{equation}

From the time evolution of the mean square displacement we then calculate the diffusion coefficient $D$  for diffusive displacements using the Stokes-Einstein equation:
\begin{equation}
\displaystyle{\lim_{t \to \infty}<r^{2}(t)>=6 D t}
\end{equation}

Finally we will use the Non Gaussian parameter $\alpha_{2}(t)$ to measure the extent of cooperative motions inside the medium.
\begin{equation}
\displaystyle{\alpha_{2}(t)=\frac{3 <r^{4}(t)>} {5<r^{2}(t)>^{2}}    -1}  \label{e06} 
\end{equation}

Unless otherwise specified the results presented in this work correspond to a temperature $T=500 K$ at which our model liquid is supercooled.
Notice however that our system is a model system with Lennard-Jones interactions only.
Therefore, due to the properties of the Lennard-Jones potential, one can shift the temperatures by any factor $\alpha$ ($T'=\alpha T$) creating a new material.
That new material will have all the $\epsilon$ parameters of the LJ potentials (including for the motor) modified in the same way $\epsilon ' = \alpha \epsilon$.
The results corresponding to that new material will have to be translated from the results of our work by a shift of time $t'=t / \alpha^{0.5}$.
Similarly, the size of the atoms can be scaled by a factor $\gamma$ ($\sigma'=\gamma \sigma$) provided that all the distances of the system are scaled by the same factor.
Our results will in that case have to be translated by a shift of distance $\gamma$ ($r'=\gamma r$) and a shift of time $\gamma$ ($t'=\gamma t$).
In that way our simulations are valid for a large number of materials although approximately, and experimentalists can adjust our data to their system of concern.

\section{Results and discussion}

\subsection{Diffusion inside the pore}

\begin{figure}
\centering
\includegraphics[height=5.65 cm]{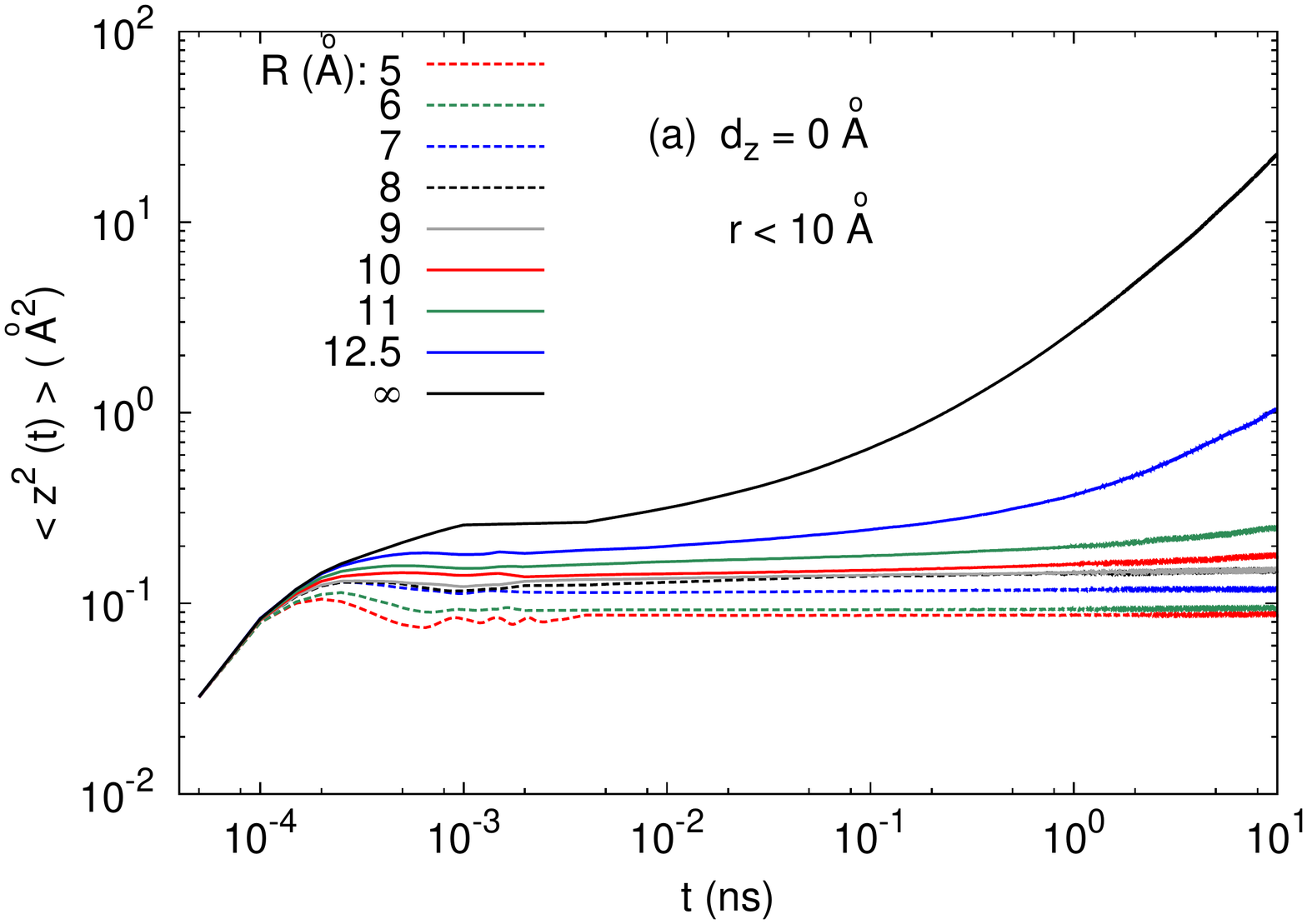}
\includegraphics[height=5.65 cm]{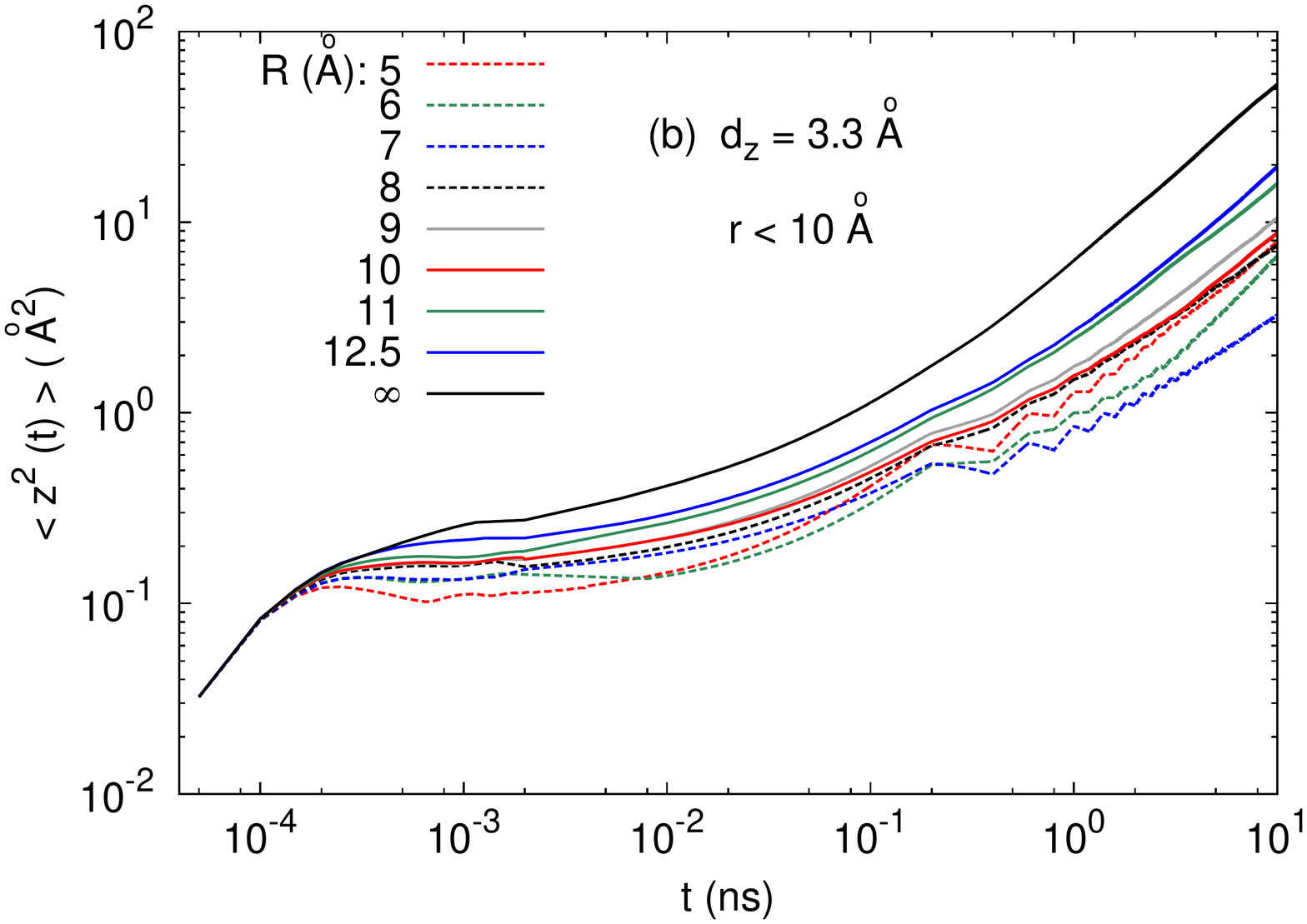}
\includegraphics[height=5.65 cm]{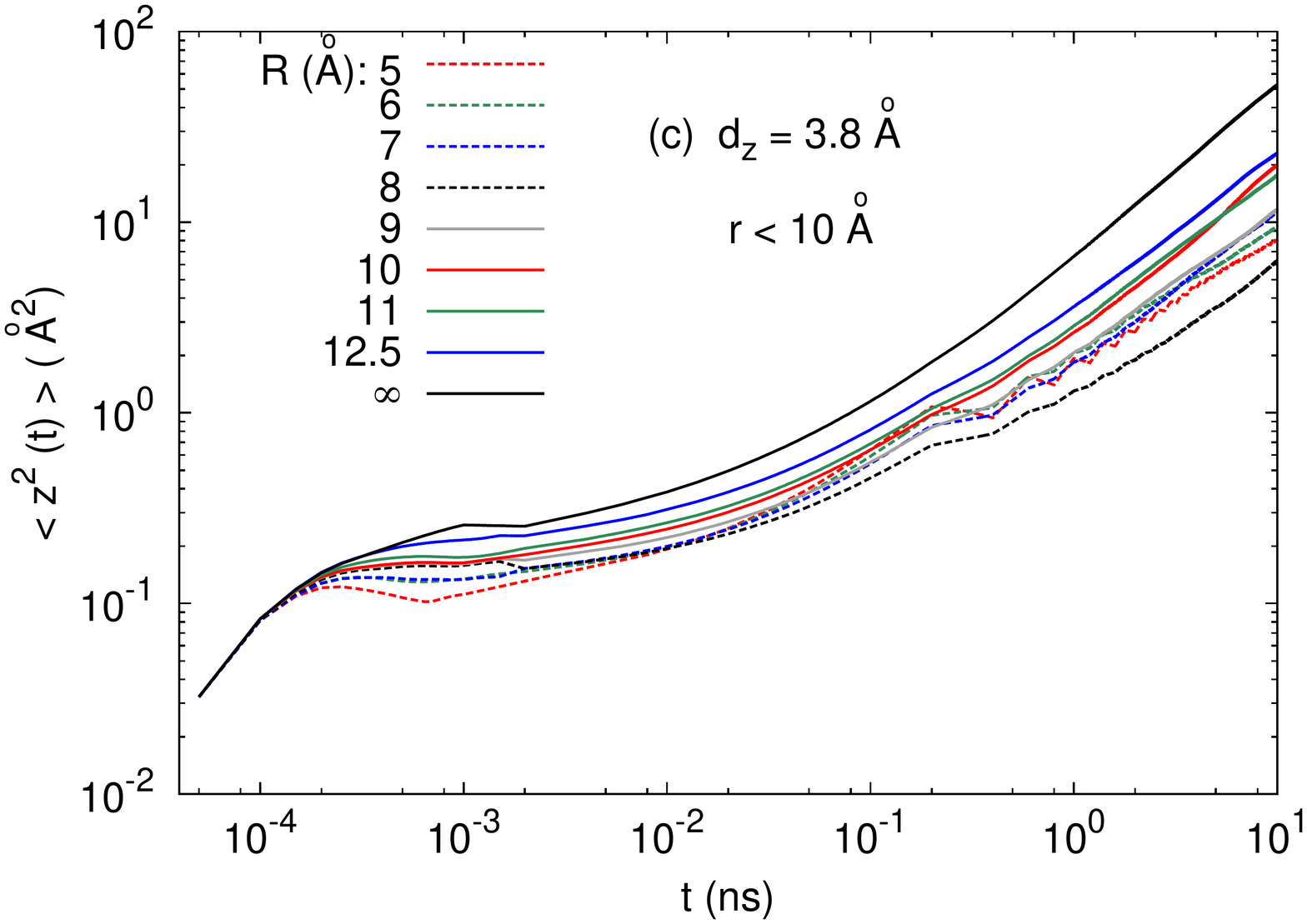}
\includegraphics[height=5.65 cm]{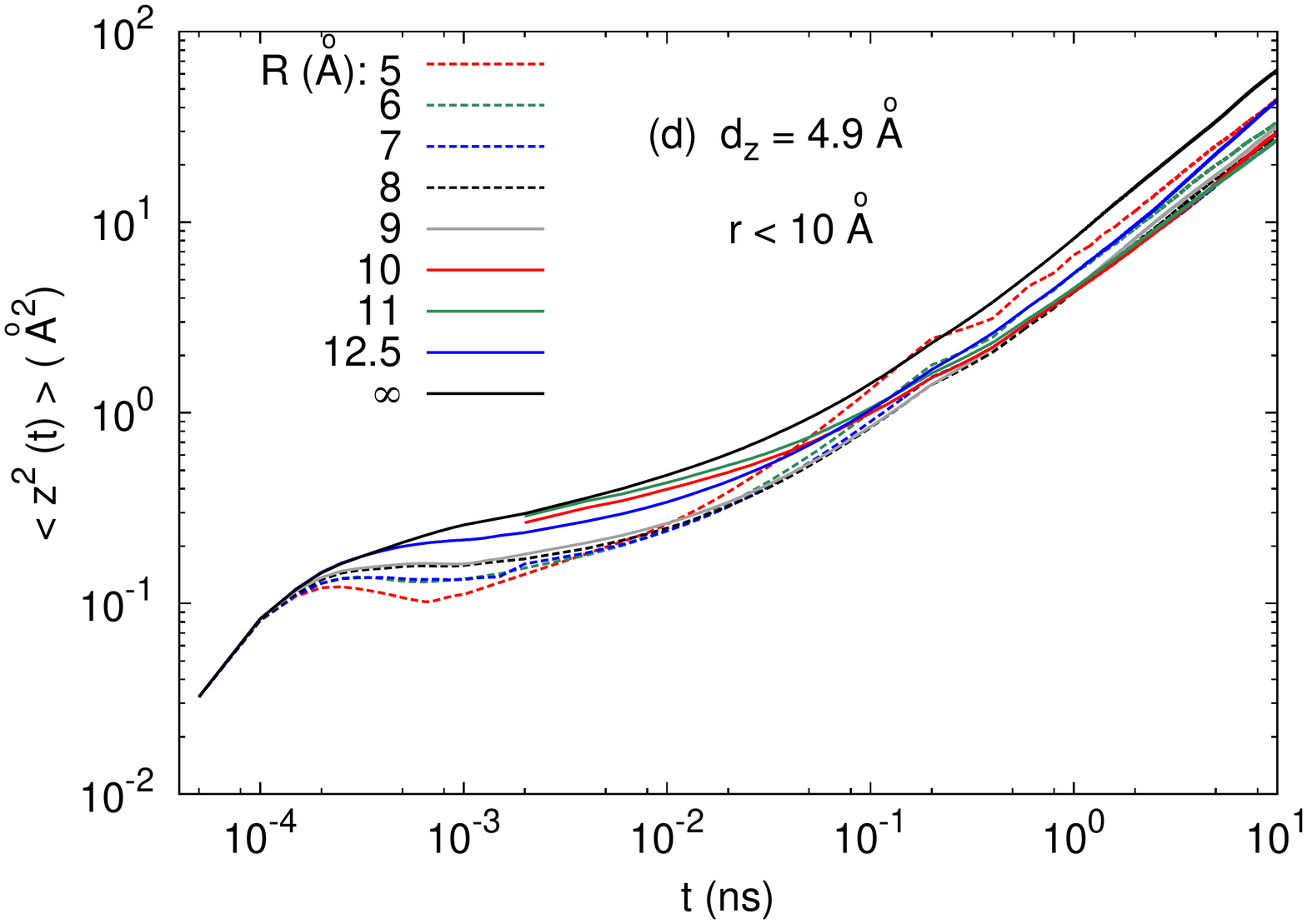}
\caption{(color online) Mean square displacement of the medium molecules located at a distance $r<10$ \AA\ from the motor at time $t=0$, for different pore radii $R$.
The activation parameter $d_{z}$ is different for each Figure. $Z$ is the direction of the pore axis and $z$ the direction perpendicular to the motor's plane in the motor's molecular frame.
In Figure (a) $d_{z}=0$ meaning that there is no activation and the motor acts as a passive probe.} 
\label{f1}
\end{figure}

Figures \ref{f1}a  \ref{f1}b  \ref{f1}c and  \ref{f1}d show the mean square displacement of the confined medium for different pore radii ranging from the Bulk to a pore radius $R=4$\AA, each Figure corresponding to a different activation quantified by the motion of the arms $d_{z}$ (Figure \ref{f0}).
For a purely passive motor molecule (Figure  \ref{f1}a), the pore effect is important. As the pore size decreases, the diffusion inside the medium decreases (i.e. the viscosity increases) leading for pores of radii smaller than 11\AA\ to a solid non diffusive medium. This confinement effect  is due to the decrease of the probability of molecules near the wall to escape the cage created by the molecules surrounding them, as part of their cage is created by the pore wall and thus never opens. Notice that, this well established confinement effect \cite{conf-1,conf0,conf1,conf2,conf3,conf4,conf5,conf6,confine,conf7}, displays a number of similarities with the effect of a decrease in temperature in supercooled liquids.
As expected, when the motor activates the medium in Figures  \ref{f1}b to  \ref{f1}d, the diffusion inside the pores increases.
However, surprisingly it increases more in the small pores (see for example the rapid increase for $R=5$\AA), than in the larger ones, leading to a decrease of the difference in diffusions.
Eventually, for the larger activation investigated in Figure  \ref{f1}d, the diffusion inside the different pores is rather similar.
Thus we observe in Figure \ref{f1} that the activation progressively suppresses the effect of confinement.

\begin{figure}[H]
\centering
\includegraphics[height=7.8 cm]{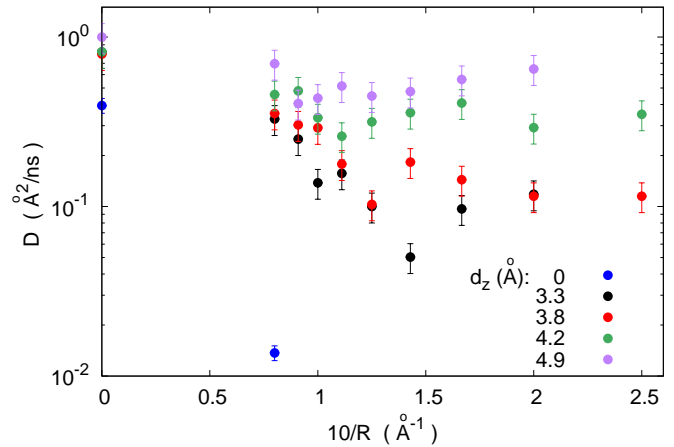}
\caption{(color online) Diffusion coefficient of the medium molecules for different pore radii $R$ and activation parameters $d_{z}$.} 
\label{f2}
\end{figure}

The evolution of the diffusion coefficient in Figure \ref{f2} resumes that behavior.
The upper circles on the Figure show the effect of the larger activation by the motor on the medium's diffusion for various pore radii.
The diffusion coefficients appear roughly constant for that large activation, even increasing slightly when the pore radius decreases.
As discussed before, the activation of the medium inside the pore  washes out the slowing down induced by the confinement.
For an activation slightly weaker the second rank of circles from the top show a constant diffusion for the different pores.
When the activation decreases the pore slowing down effect increases but is still much smaller in the Figure than for the non-activated medium ($d_{z}=0$).

\subsection{Activation effect}

Figure \ref{f2}  shows that the diffusion coefficients follow exponentially decreasing laws with the inverse of the pore radius $1/R$.

\begin{equation}
\displaystyle{D= D_{0} exp(-{\zeta}/R)}  \label{e00} 
\end{equation}

Where $\zeta$ is a correlation length that depends on the activation parameter $d_{z}$.
As a result, one can explain the decrease of the pore effect with the activation, by the decrease of that correlation length of the medium when it is activated.
For our largest activation the correlation length $\zeta$ tends to zero and the pore effect disappears.

\begin{figure}
\centering
\includegraphics[height=6.8 cm]{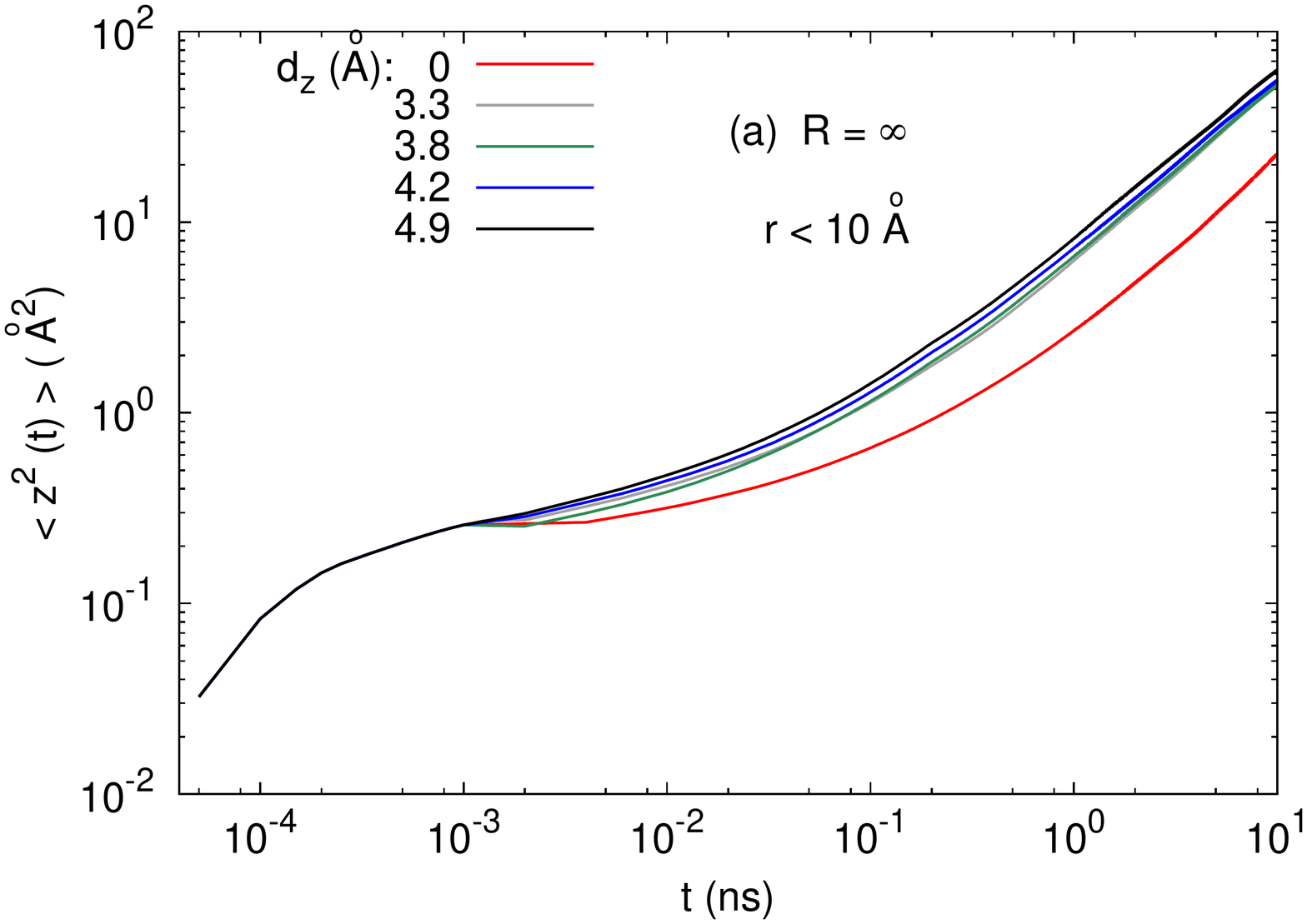}
\includegraphics[height=6.8 cm]{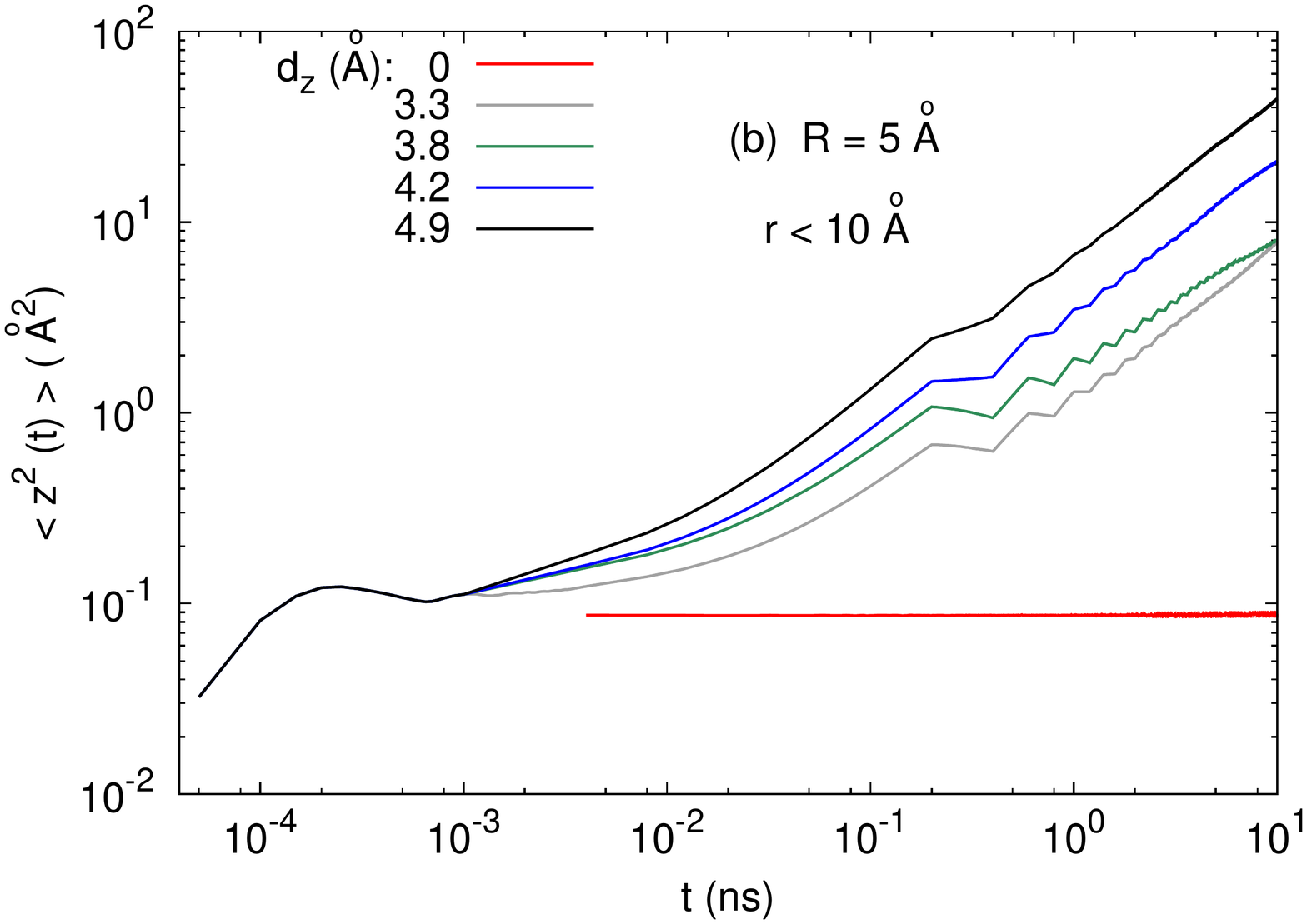}
\caption{(color online) Mean square displacement of the medium molecules for various activation parameters $d_{z}$; (a) in the bulk ($R=\infty$); (b) for a pore of radius $R=5$\AA.
The activation dependence appears much larger in the pore.} 
\label{f3}
\end{figure}

Figures \ref{f3}(a) and (b) compare the effect of the activation in the bulk and in a pore of radius $R=5$\AA.
Figure \ref{f3} shows a much larger activation effect on the medium's displacements inside the small pore than in the bulk.
For a time lapse of $10 ns$ the mean square displacement in the pore varies indeed from $10^{-1}$\AA$^{2}$ to $4.10^{1}$\AA$^{2}$, that is by a factor $400$ depending on the activation, while in the bulk it varies by a mere factor $3$.
This result agrees well with the findings of previous section that the diffusion for large activation doesn't depend on the pore size.
As the pore radius decreases, the slowing down by the pore wall increases, but the activation effect also increases compensating that slowing down.
Notice that as the pore radius decreases, the relative number of motor to free molecules increases due to the geometry.
While results we show take only into account molecules a distance $r<10$\AA\ apart from the motor, reducing that geometrical effect, it doesn't eliminate it.
Geometry appears consequently as a possible cause for the increase of the activation impact on the diffusion for small pores.
In order to conclude on that possibility we calculated the mean square displacements at a smaller distance $r<5$\AA\ from the motor.
The results (not shown) were very similar to the results of Figure \ref{f3}, showing that the increase of the activation effect on diffusion for small pores, is not arising from the geometry (i.e. from the mean distance to the motor). 

It is interesting to point out however that the increase of the effect of activation when the pore size decreases doesn't increase the diffusion to values larger than  bulk values but tends instead to reach the bulk values from below.
That point comes in support of the decrease of the correlation lengths explanation. 

\subsection{Cooperative motions}

\begin{figure}
\centering
\includegraphics[height=7.5 cm]{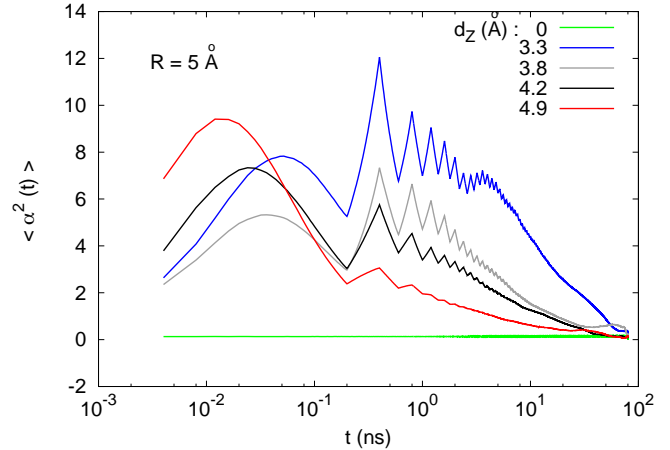}

\caption{(color online) Non Gaussian parameter $\alpha_{2}(t)=\frac{3 <r^{4}(t)>} {5<r^{2}(t)>^{2}}    -1$ around the motor ($r<10$\AA) for various activation parameters $d_{z}$ and for a pore of radius $R=5$\AA, at $T=500 K$.} 
\label{f10}
\end{figure}

Spontaneous (thermal)\cite{dh0} and activation-induced  cooperative motions\cite{md16,c3,c4} are present in supercooled liquids and have been well documented.
Because cooperativity is directly related to the physical correlation length of the liquid, its increase or decrease can give us information on that correlation length. 
Therefore, to shed some light on the correlation length explanation, we now study the extent of cooperative motions in our system.
For that purpose we display in Figure \ref{f10} the Non Gaussian parameter $\alpha_{2}(t)$ evolution with the activation.
Cooperative motions in supercooled liquids induce a tail in the Van Hove self correlation function, leading to its departure from the Gaussian shape.
For that reason the departure from the Gaussian shape of the Van Hove, measured by the Non Gaussian coefficient, has been used extensively as a measure of the cooperative motions.

In Figure \ref{f10} on the right  $\alpha_{2}(t)$ shows peaks at times corresponding to multiple of the motor's folding period, that are the result of the excitations induced by the motor's folding on the medium.
The activation-induced cooperativity as measured by $\alpha_{2}(t)$ decreases when we increase the activation parameter $d_{z}$ from $3.3$\AA\ to $4.9$\AA.
That result comes again in support of the decrease of the medium's correlation length with the activation. 
Notice however that as expected when the activation parameter is null the induced cooperativity disappears as there is no more activation.
This result suggests an increase of the cooperativity for small activations and an optimum.

\subsection{Excitation density}

\begin{figure}
\centering
\includegraphics[height=5.6 cm]{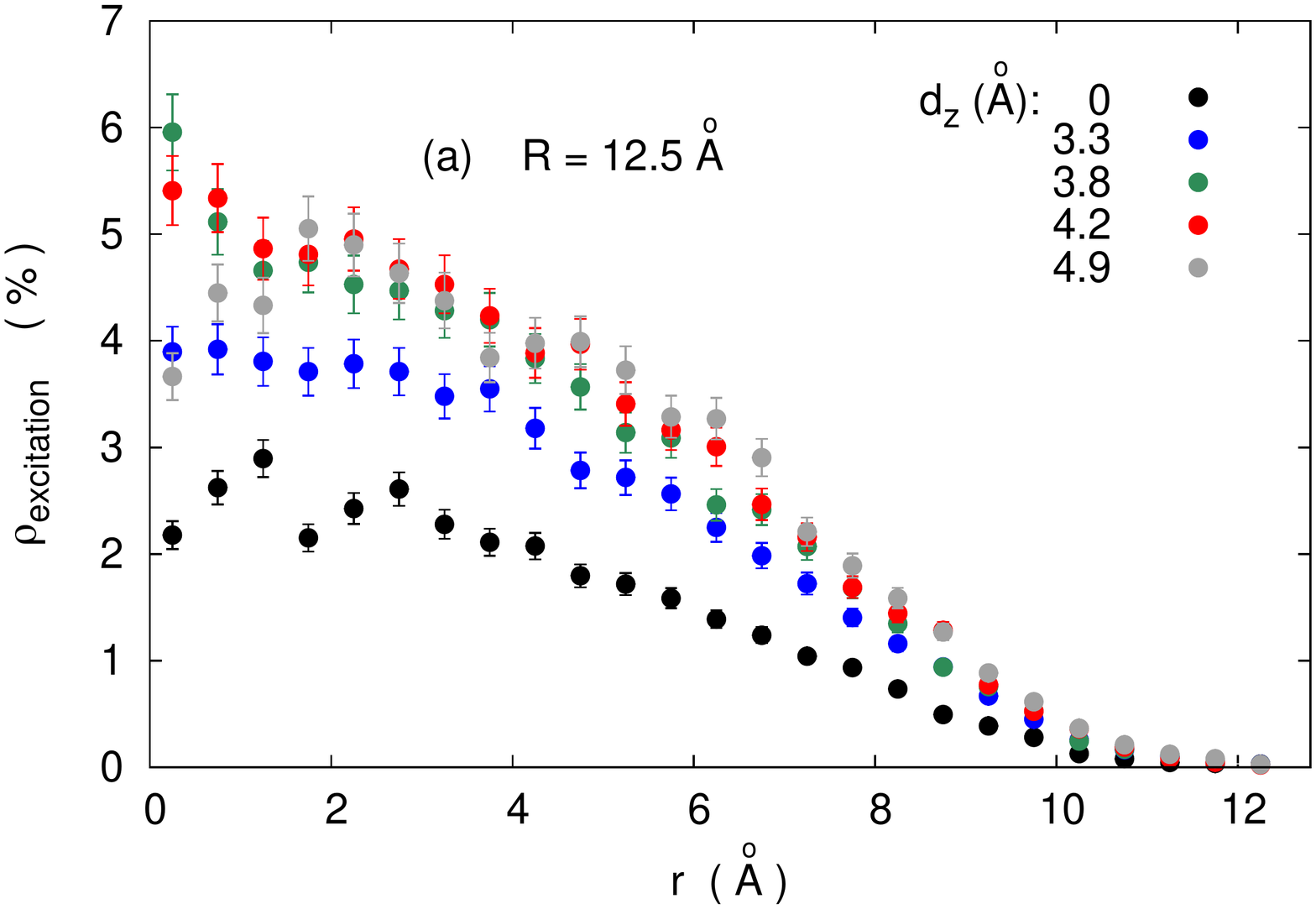}
\includegraphics[height=5.6 cm]{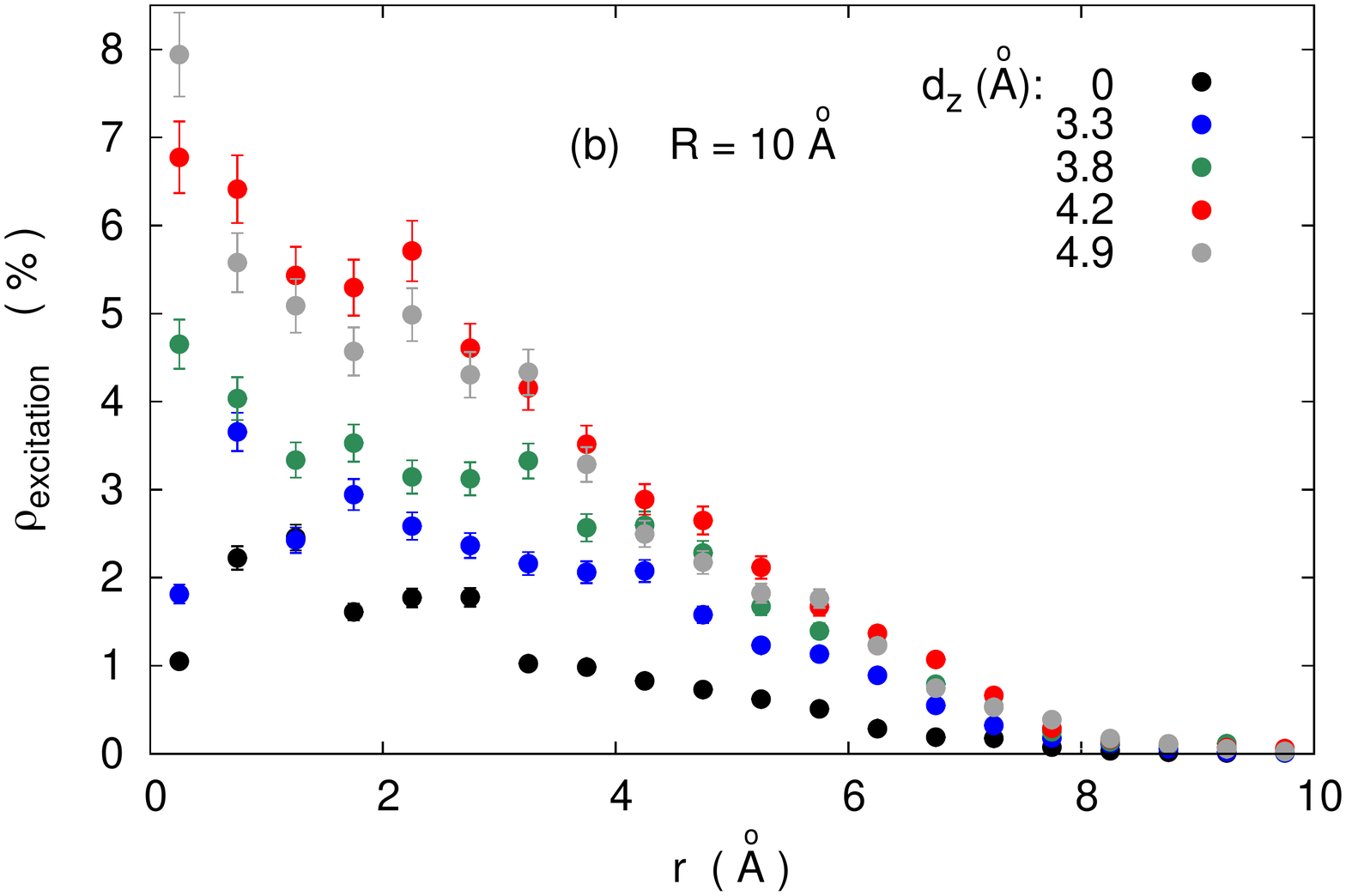}
\includegraphics[height=5.6 cm]{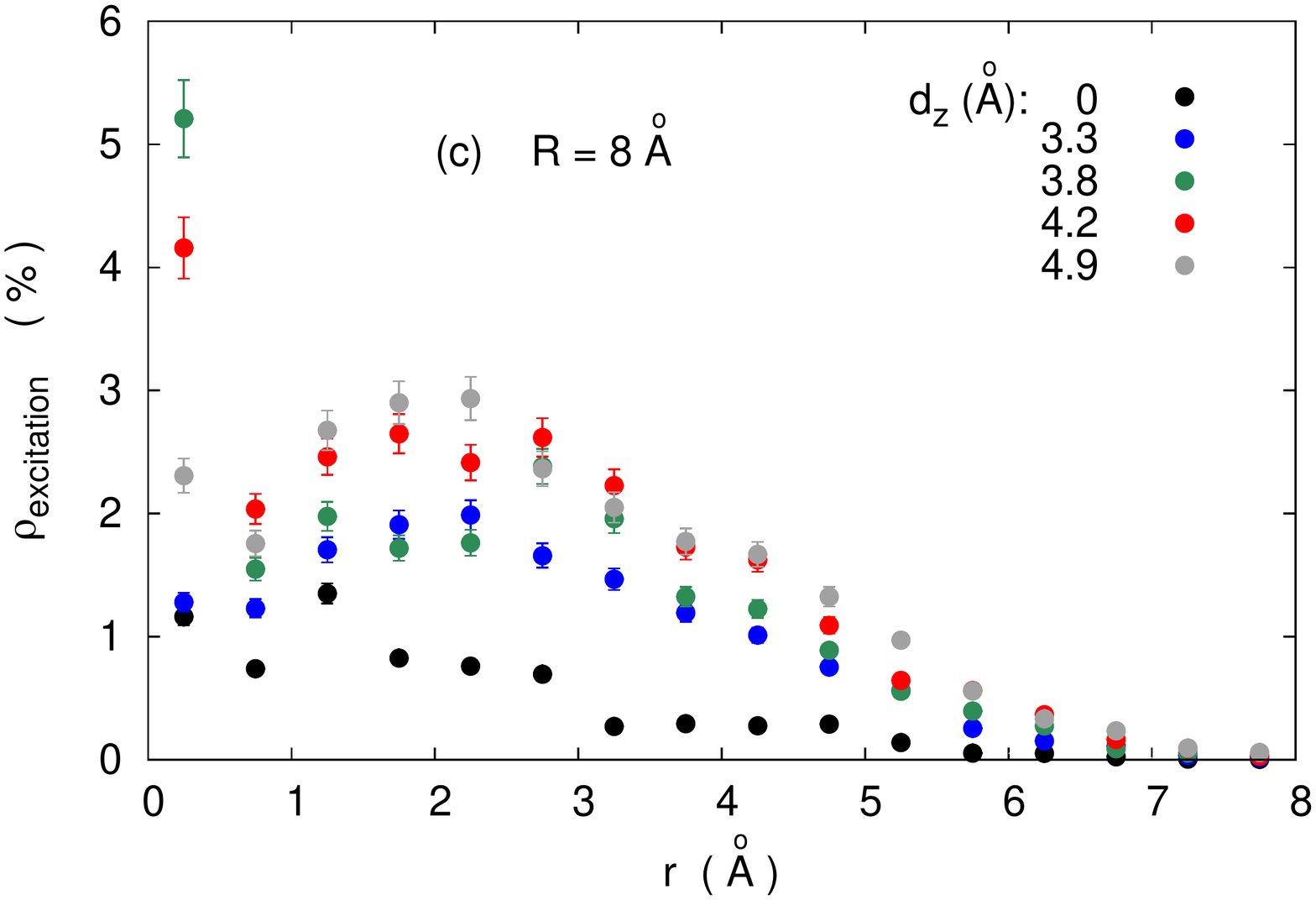}
\includegraphics[height=5.6 cm]{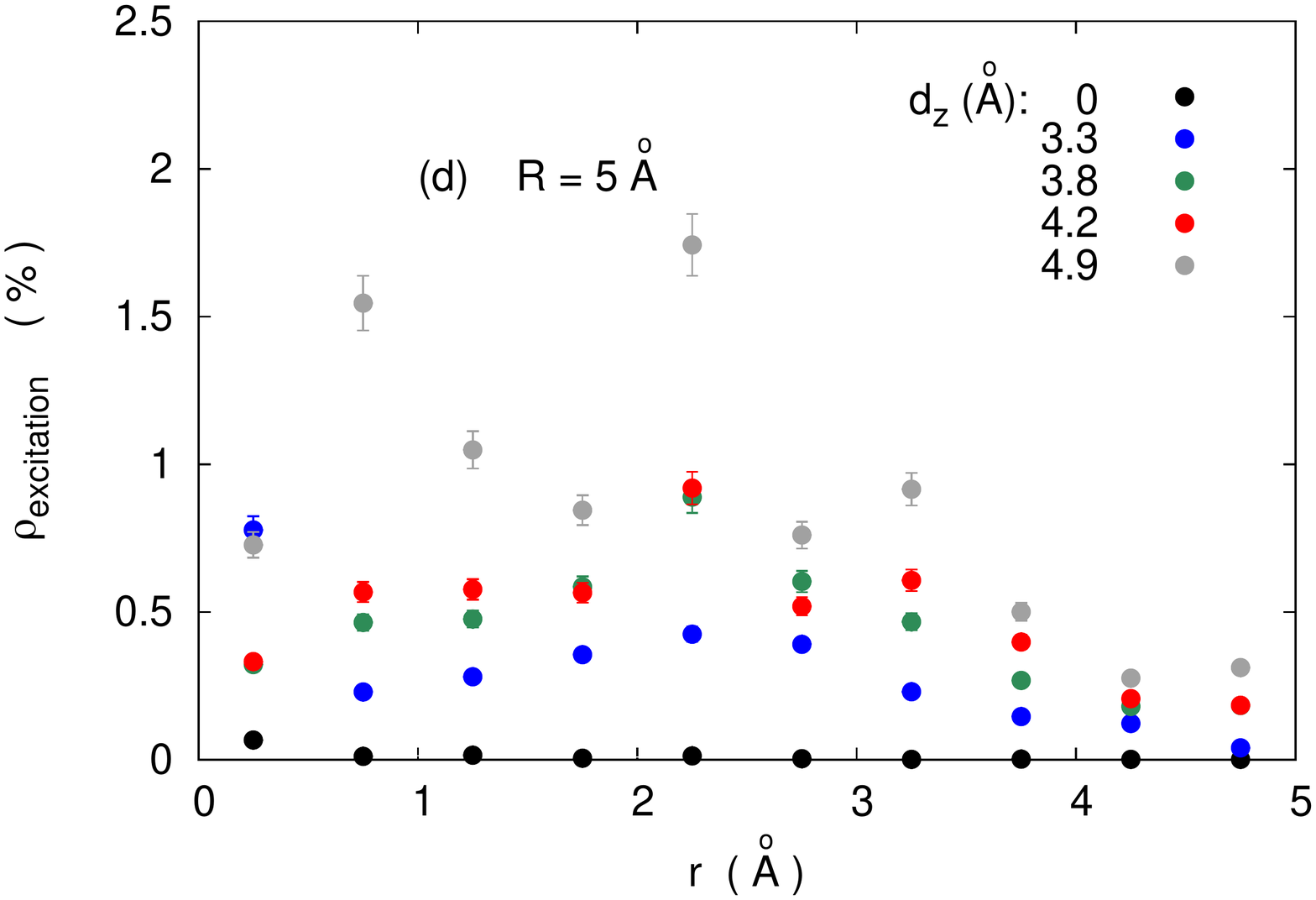}
\caption{(color online) Excitation density $\rho_{excitation}$ as a function of the distance $r$ from the pore central axis, for different pore radii $R$ and activation parameters $d_{z}$. Excitations are here defined as medium's molecules motion larger than $\Delta r=1$ \AA\ occuring during a time lapse $\Delta t=10 ps$.} 
\label{f4}
\end{figure}

To better understand previous results, we now turn our attention to the density of excitations inside the pores.
Following Keys et al. \cite{keys} we define excitations as elementary diffusion processes. In our work, excitations are molecules that move in a time lapse $\Delta t=10 ps$  a  distance larger  than $\Delta r=1$ \AA\ from their previous position. Figure \ref{f4} shows the excitation density for various pore radii and excitation processes.
We find in Figure \ref{f4} that the density of excitation increases with the activation whatever the pore radius.
That result validates at posteriori our choice of the activation parameter as the displacement of the motor's arm $d_{z}$.

Interestingly  the maximum excitation density is not always located at the pore center.
This effect could be due to layering inside the pores.
Layering induces oscillations of the density inside the pores and low densities promote excitations as cages are then loose, letting molecules escape easier.
We will test that interpretation in Figure \ref{f6}.

Figure \ref{f4} shows that when the pore size decreases the excitation density decreases much less rapidly for large activations.
Therefore the excitation density is larger than expected for small pores when the activation is on, in agreement with the activation effect described in a previous section.
That result suggests that the larger than expected diffusion observed for small pores (the activation effect) is due to a larger density of excitations that is a larger number of mobile molecules as opposed to larger motions of some molecules.

\subsection{Structural organization}

\begin{figure}
\centering
\includegraphics[height=5.6 cm]{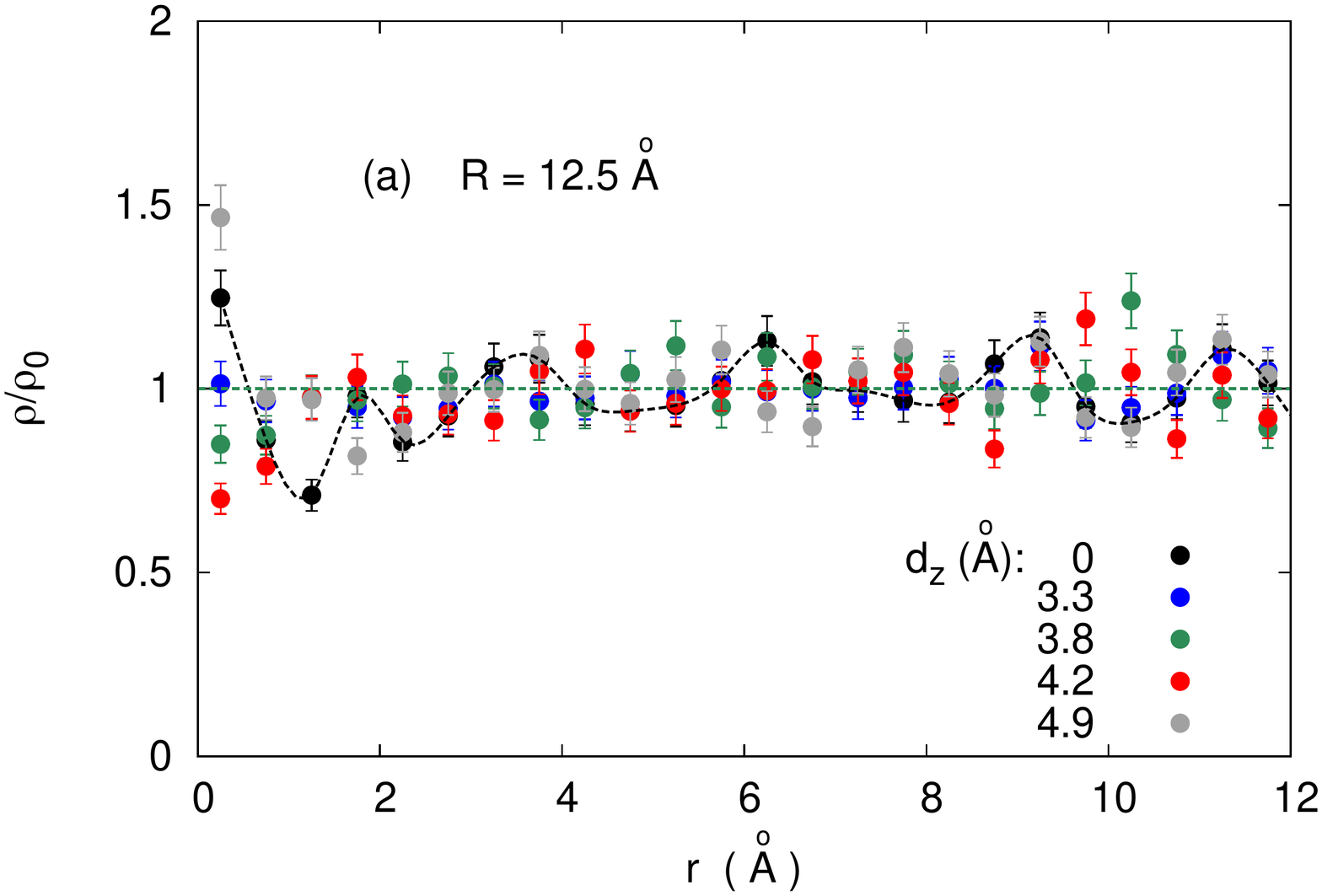}
\includegraphics[height=5.6 cm]{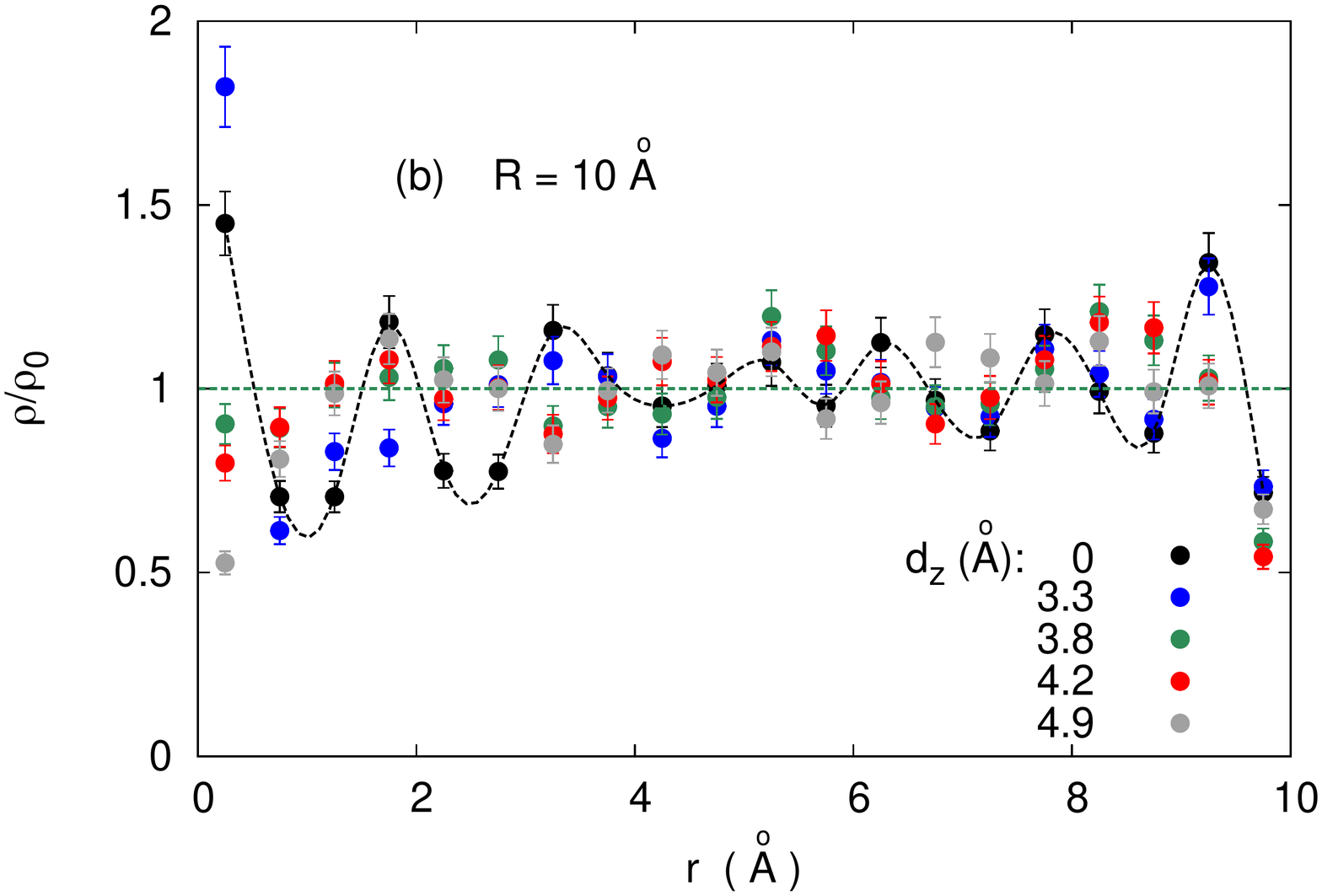}
\includegraphics[height=5.6 cm]{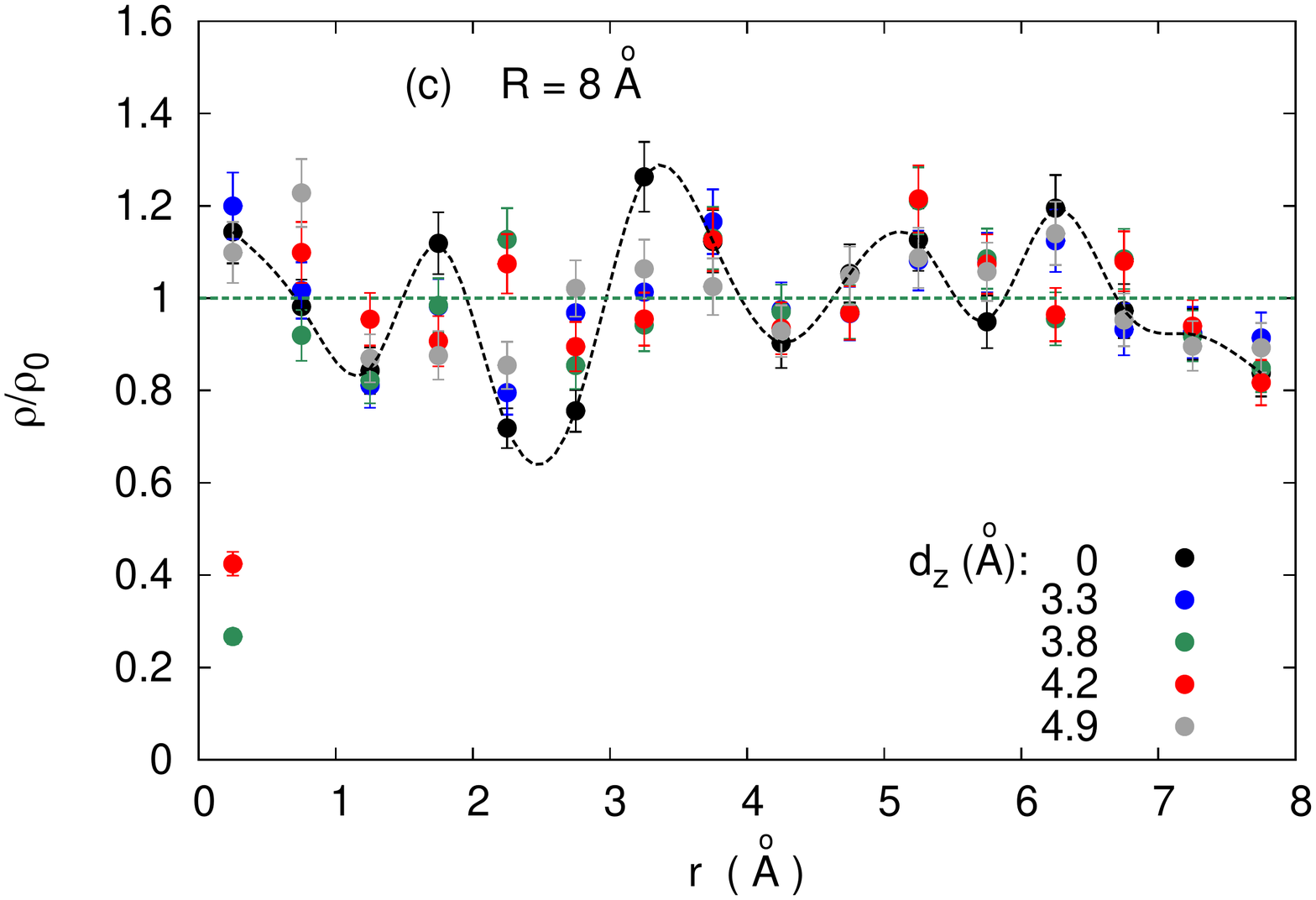}
\includegraphics[height=5.6 cm]{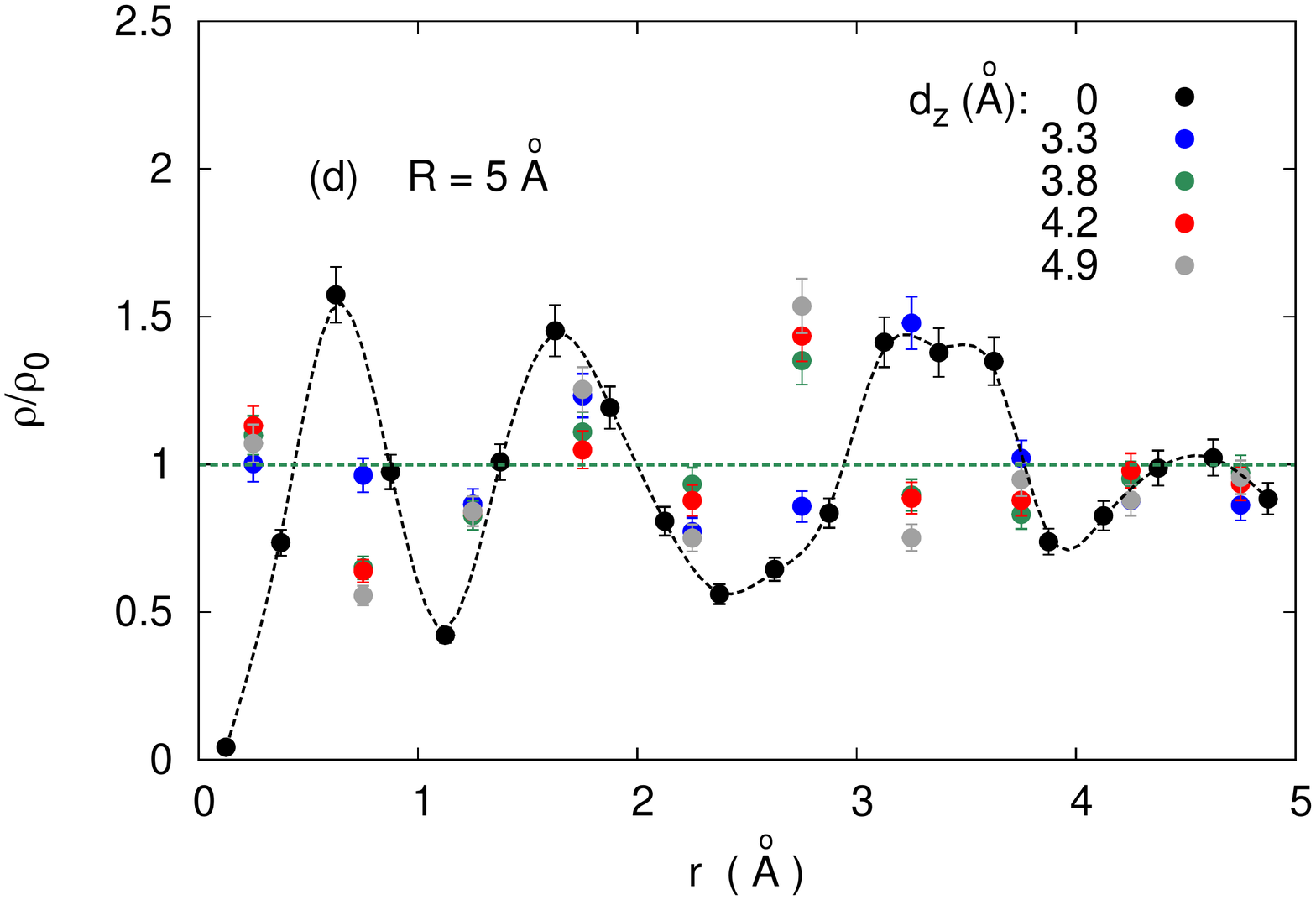}
\caption{(color online) Density  $\rho$ of medium molecules inside the pore as a function of the distance $r$ from the pore central axis. The Figures show that result for different pore radii $R$ and activation parameters $d_{z}$. The values are normalized by the average density inside the pore $\rho_{0}$. The dashed line displays the layering when the motor is not active.
The figures show that the layering is strongly modified with the activation.} 
\label{f6}
\end{figure}

We will now investigate the local structure inside the pores.
In Figure \ref{f6} we display the local density inside the pores as a function of the distance $r$ from the pore center.
 That is, $r$ is the radius of cylindrical coordinates, $z$ being the pore axis.
The Figure shows a clear layering effect, that we highlight with a line when the activation is off.
The local density is in most cases as expected maximum at the center of the pore ($r=0$) and minimum around the wall.
However this is not always the case. For instance for a pore of $R=5$\AA\ radius, the maximum density is not located in the center of the pore, and around the wall for large activations the density is not null.
For a pore of $R=8$\AA\ radius, there is a maximum density at the pore center but it corresponds to a thin peak, and we observe a wider maximum apart from the center.

As expected, the layering explains the excitation density maxima shifted from the center of the pore, as the local density is actually smaller at the location of excitation density maxima.
Moreover an unexpected effect appears,  the layering changes with the activation. 
 Not only it decreases slightly but the oscillations locations are also modified.
 Work is in progress to better understand that interesting effect.

\section{Conclusion}
In supercooled liquids, nano-confinement modifies the dynamical properties of the liquid usually inducing an important  slowing down of the dynamics.
Confinement also induces a structural organization called layering. 
In this work we investigated how these confinement effects are modified when the liquid is out of equilibrium.
Results show that  the dynamical slowing down induced by the confinement is strongly modified by the activation, leading to a diffusion that may not depend anymore on the pore diameter.
In agreement with that finding our results show that the activation effect is  much larger upon confinement than in the bulk, increasing rapidly when the pore size decreases.
We interpret these effects as arising from an activation induced decrease of the dynamical correlation lengths of the liquid.
In agreement with that picture we observe a decrease of the cooperative motions upon activation.
While the structure of the liquid is unchanged, we observe a decrease of the layering inside the pore but also a qualitative modification of the layering with the activation.
Eventually we observe a relation between the layering and the elementary diffusion processes called excitations, a result strongly suggesting that the layering triggers an increase of the mobility in the less dense parts of the pore.

\vskip 0.5cm

{ \bf \large  Conflict of interest}

There are no conflict of interest to declare.
\vskip 0.5cm

{ \bf \large  Data availability}

The data that support the findings of this study are available from the corresponding author upon reasonable request.
\vskip 0.5cm


\end{document}